\documentclass[11pt,a4paper]{article}

\usepackage[utf8]{inputenc}
\usepackage[T1]{fontenc}
\usepackage{lmodern}
\usepackage[margin=2.6cm]{geometry}
\usepackage{amsmath,amssymb,amsthm}
\usepackage{booktabs,tabularx,array}
\usepackage{enumitem}
\usepackage{float}
\usepackage{tikz}
\usetikzlibrary{arrows.meta,positioning}
\usepackage{xurl}
\usepackage[hidelinks]{hyperref}
\usepackage{microtype}

\setlist{itemsep=2pt,topsep=4pt}
\newcolumntype{L}{>{\raggedright\arraybackslash}X}
\newcolumntype{P}[1]{>{\raggedright\arraybackslash}p{#1}}

\theoremstyle{definition}
\newtheorem{definition}{Definition}
\newtheorem{assumption}{Assumption}
\theoremstyle{plain}
\newtheorem{claim}{Claim}

\newcommand{\KDF}{\mathsf{KDF}}
\newcommand{\Hash}{\mathsf{H}}
\newcommand{\Gen}{\mathsf{Gen}}
\newcommand{\Rep}{\mathsf{Rep}}
\newcommand{\AEnc}{\mathsf{AE}}
\newcommand{\concat}{\,\|\,}

\title{Baryonic Keys and the Inheritance of Crypto-Assets:\\
Revisiting the Tales From the Crypt Protocol}
\author{Fr\'ed\'eric Prost\\[2pt]
\small INSA Lyon, INRIA, CITI, UR3720, Villeurbanne, France\\
\small \texttt{frederic.prost@insa-lyon.fr}}
\date{November 2026}

\begin{document}
\maketitle

\begin{abstract}
The inheritance of crypto-assets faces a structural obstacle: ownership is the
knowledge of a key, so the key either disappears with its owner or has to be
shared beforehand. The Tales From the Crypt Protocol (TFCP) addressed one part
of the problem, the distributed and privacy-preserving acknowledgment of a death
by a blockchain. We revisit it with a new primitive. A \emph{baryonic key} is a
physical object from which bits are obtained only through a measurement, are
never stored digitally, and which exists in exactly $n$ copies, where $n$ is
neither zero nor unbounded. We define the primitive, outline a candidate
instantiation based on a replicated micro-fractured relief, and state its
limits: digitising a copy is slow and tamper-evident, not impossible. We also
outline other uses of the primitive, from post-quantum pre-shared keys to common
coins for Byzantine agreement. We then give a threat model for TFCP, review the
attack models of the literature that apply to it, and show where it resists and
where it does not. We propose TFCP-B, in which decrypting the will requires both
a secret shared among registrars and the response of a baryonic key, and in
which heirs claim their share through hash commitments, without any prior key or
wallet. We detail each change from TFCP to TFCP-B, with its rationale and its
price. TFCP-B removes the anonymity loss caused by colluding registrars and
blocks several false-alarm attacks. It does not remove the shallow-account
attack, remains exposed to transaction censorship during the deliberation
period, and introduces new assumptions on custodians and reading devices. The
security arguments are informal and the physical figures are design estimates.
\end{abstract}


%=====================================================================
\section{Introduction}

A crypto-asset is controlled by whoever knows the corresponding private key.
This is the source of its resistance to seizure and censorship, and also the
source of a difficulty that is seldom addressed: when the owner dies, either the
knowledge of the key dies as well, or the key has been given to somebody else in
advance, in which case ownership was shared during the owner's
lifetime~\cite{p1,p2}.

Two earlier papers explored this difficulty. The first one~\cite{p1} proposed
the Tales From the Crypt Protocol (TFCP), by which a blockchain can acknowledge
the death of an account owner without a trusted oracle and without revealing,
before the death, which account belongs to whom. It deliberately left aside the
delivery of credentials to the heirs. The second one~\cite{p2} listed open
questions, and suggested in its Section~4.2 that secret shares could be
``objectified'': turned into physical objects cloned by an analogue process,
from which shared bits would later be obtained by measurements.

The observation behind that suggestion is that digital information admits only
two regimes of copying. A secret known to a single person has no copy and
disappears with that person. A secret written once in a computer memory can be
copied perfectly, silently and without bound. Inheritance requires an
intermediate regime: a secret that exists in exactly $n$ places, with $n$ small
and fixed, and that cannot be duplicated in a second.

\paragraph{Contributions.}
\begin{enumerate}
  \item We define \emph{baryonic keys}, physical keys that are clonable a
bounded number of times, and outline a candidate instantiation
(Section~\ref{sec:keys}). We position the primitive with respect to physical
unclonable functions and to recent work on twin and distributable physical
functions (Section~\ref{sec:related}), and outline its uses beyond inheritance
(Section~\ref{sec:uses}).  \item We give a system and threat model for
death-acknowledgment protocols (Section~\ref{sec:model}).
  \item We specify TFCP-B, a revision of TFCP built on baryonic keys
(Section~\ref{sec:tfcpb}). The will is encrypted under a key derived from two
halves that are useless in isolation; heirs claim \emph{slots} defined by hash
commitments. We then make explicit what changes with respect to TFCP, why, and
at what price (Section~\ref{sec:evolution}).  \item We analyse TFCP and TFCP-B
against attack models taken from the literature on bribery, censorship,
front-running, flash loans, fuzzy extractors and physical unclonable functions
(Section~\ref{sec:analysis}), and state for each of them whether the protocols
resist.
  \item We revisit the open questions of~\cite{p1,p2}
(Section~\ref{sec:discussion}), compare costs with those of existing services
(Section~\ref{sec:costs}) and list the limitations of the approach
(Section~\ref{sec:limits}).  \end{enumerate}

\paragraph{Scope.} The security arguments of this paper are informal. The
physical figures are design estimates that have not been validated
experimentally. Both points are discussed in Section~\ref{sec:limits}.

%=====================================================================
\section{Background and related work}
\label{sec:related}

\subsection{The Tales From the Crypt Protocol}
\label{sec:tfcp}

We keep the vocabulary of~\cite{p1}. The \emph{Donor} is the account of the
person whose death has to be acknowledged. The \emph{Security Deposit} is an
account controlled by the Donor and used to signal the death. \emph{Witnesses}
testify to the death by transferring an \emph{Ante} to the Security Deposit.
\emph{Registrars} stake a \emph{Bail} and hold \emph{Shares} of a key that
encrypts the \emph{Wills}, a document linking the Donor's account to a public
\emph{Announcement} that bears the civil identity of the Donor.

The Donor recruits Registrars, hands them the Shares, and publishes the
encrypted Wills and the Announcement. After the death, Witnesses transfer their
Antes. Once a \emph{Threshold} is reached, a \emph{Deliberation Time} starts. If
the Donor's account moves during that time, the Donor is alive: the process
stops and the Security Deposit, which now holds the Antes, becomes spendable by
the Donor. Otherwise the Registrars decrypt the Wills, the \emph{Acknowledgment}
is enacted by the network, fees are paid to Registrars and Witnesses, and
inheritance transactions are executed. The content of these transactions is
outside the scope of~\cite{p1}.

The stated properties are a distributed oracle, the anonymity of the Donor's
account before the Acknowledgment under an honest-but-curious assumption on
Registrars, and independence from the underlying blockchain. Two attacks are
discussed in~\cite{p1}: a Donor who hands the key of an almost empty account to
an accomplice in order to capture the Antes, and a wealthy Witness who reaches
the Threshold alone.

\subsection{Inheritance of crypto-assets in practice}

Deployed solutions have converged on the dead man's switch, which~\cite{p2}
already criticised for the constant vigilance it requires. Liana adds to the
owner's key a recovery key that only becomes valid after a period of inactivity,
the delay being enforced by the Bitcoin network itself~\cite{liana}; Nunchuk
launched an inheritance scheme based on the same on-chain mechanism at the end
of 2025~\cite{spark}. Collaborative custody services such as Casa rely on
multi-signature wallets in which the provider holds one key~\cite{casa}.
Sarcophagus is a dead man's switch operated by staked third
parties~\cite{sarco}. A legal analysis of crypto-asset succession in civil-law
systems is given in~\cite{carata}.

These solutions share two limits. First, the trigger is the owner's silence
rather than the owner's death, so that a short delay may open the succession of
a person who is travelling or in hospital, while a long delay postpones every
legitimate succession. Second, the heir has to hold a key or a device as of
today. Neither the distributed acknowledgment of a death nor the transmission of
rights to a party that is not yet known is addressed.

\subsection{Physical unclonable functions and shared physical keys}

Physical one-way functions were introduced by Pappu et al.~\cite{pappu}, who
used the speckle pattern of a disordered optical token as an identifier that is
cheap to make and very hard to duplicate. The large family of physical
unclonable functions (PUFs) that followed is surveyed in~\cite{mcgrath}. A PUF
is unique by design: the party that holds it must extract a key and transmit it
by conventional means if another party is to share it.

Two recent lines of work depart from uniqueness. Zhong et al.~\cite{zhong}
fabricate \emph{twin} PUFs on the same aligned carbon nanotube array; the two
devices agree on about 95\,\% of their bits and are used as a shared key that is
never extracted or stored. Luescher et al.~\cite{luescher} propose
\emph{chemical} unclonable functions based on random DNA pools, which are
explicitly designed to be distributable: a pool can be split into identical
pools held by several parties. A holder who knows the structure of the pool can
amplify it further, and the authors add a chemical modification to disable
copying.

Baryonic keys belong to this second family. They differ in the property that is
taken as primary: the number of copies is fixed once and for all by the
destruction of the replication master, and the design goal is the transmission
of secrets across decades rather than authentication or communication.

Reconciling noisy physical readings into identical keys is the purpose of fuzzy
extractors~\cite{drs}. Two refinements matter here: reusability, since several
enrolments of the same source may leak more than a single one~\cite{boyen}, and
robustness against an adversary who modifies the public helper data~\cite{dkrs}.
Modelling attacks, in which a machine-learning model trained on observed
challenge--response pairs predicts new responses, have broken several electrical
PUF designs~\cite{ruhrmair}.

%=====================================================================
\section{Baryonic keys}
\label{sec:keys}

\subsection{Definition}

\begin{definition}[Baryonic key]
A baryonic key with parameters $(n, \ell, \tau)$ consists of $n$ physical
objects, the \emph{copies}, a challenge space $\mathcal{C}$, and a public
reading procedure which, on a copy and a challenge $c \in \mathcal{C}$, returns
a noisy string $w(c)$. Together with a fuzzy extractor $(\Gen, \Rep)$, it
defines a response $R(c) \in \{0,1\}^{\ell}$ and public helper data $P(c)$ such
that the following properties hold.
\begin{description}[style=nextline,leftmargin=1.2cm]
  \item[B1 -- Agreement.] For every challenge $c$ and every copy, $\Rep$ applied
to a reading of that copy and to $P(c)$ returns $R(c)$, except with negligible
probability.
  \item[B2 -- No digital storage.] No representation of the responses exists
outside the matter of the copies; $P(c)$ reveals a negligible
amount of information on $R(c)$.
  \item[B3 -- Bounded clonability.] Exactly $n$ copies exist. Producing a
further copy, or a digital model that predicts responses to arbitrary
challenges, requires physical access to a copy for a time at least $\tau$ and
leaves physical evidence.  
  \item[B4 -- Independence.] Knowledge of the responses to a set of challenges
gives a negligible advantage in predicting the response
to a challenge whose reading does not overlap theirs.
  \item[B5 -- Durability.] The copies are passive and remain readable for decades.
\end{description}
\end{definition}

A PUF corresponds to the case $n = 1$ with $\tau$ ideally infinite. The twin
PUFs of~\cite{zhong} correspond to $n = 2$. The pools of~\cite{luescher} are
distributable, with a number of copies that the holders themselves can increase
unless copying is chemically disabled. A baryonic key fixes $n$ at fabrication
and accepts a finite $\tau$.

\paragraph{On the name.} The word \emph{baryon} was introduced by Pais in
1953~\cite{pais}, from the Greek \emph{bar\'ys}, heavy, for the family of heavy
particles to which the proton and the neutron belong. By extension, baryonic
matter is the ordinary matter of which atoms, and therefore all everyday
objects, are made; cosmologists use the term to set it apart from what is not
made of atoms. We borrow the adjective for a similar contrast. A digital key is
information: it has no weight and no location, and it is duplicated at no cost.
A baryonic key is a piece of matter: it weighs something, it is somewhere, and
its bits do not exist as information until they are measured.

\subsection{A candidate instantiation}
\label{sec:disc}

We outline the instantiation that we are studying. All figures in this
subsection are design estimates.

\paragraph{Object.} A glass master is roughened by micro-fracturing (fine
abrasive blasting followed by a mild etch), which produces a random relief with
a roughness of 50--200\,nm and a correlation length of about 1\,$\mu$m. The
relief is replicated $n$ times by direct nanoimprint~\cite{chou} onto glass
discs of 120\,mm in diameter. Each copy is coated with a mirror and sealed under
a thin transparent layer. The master is then destroyed. Replication is physical
and parallel, so that no file describing the relief exists at any time.

\paragraph{Reading.} Optical pickup units designed for optical discs measure a
focus-error signal with nanometre sensitivity, and have been used as low-cost
profilometers and as sensing heads for atomic force
microscopes~\cite{hwu,sekar}. The reader uses such a unit to measure the relief
at a few sites of $10 \times 10$\,$\mu$m. A challenge is expanded by a hash
function into a list of sites; the mapping is public.

\paragraph{Agreement.} Two copies read by two devices are expected to differ on
a few per cent of the raw bits. $\Gen$ is run once, on one copy, and outputs
$R(c)$ and $P(c)$; $\Rep$ is then run on any copy. Each site is used by a single
run of $\Gen$, which avoids the reusability issue of~\cite{boyen}.

\begin{table}[H]
\centering
\small
\begin{tabularx}{\linewidth}{@{}P{6.6cm}L@{}}
\toprule
Valid measurement sites per side & $\approx 8 \times 10^{7}$\\
Net bits per site, at 2\,\% raw error & $\approx 7.4$\\
256-bit response, 2\,\% raw error & 50 sites, read in $\approx 18$\,s\\
256-bit response, 8\,\% raw error (ageing margin) & 122 sites, read in $\approx 35$\,s\\
Number of copies $n$ & 2 to 5\\
Digitisation of one side, $\tau$ & 1--10\,h with laboratory equipment; 10--100\,h with portable equipment\\
Fabrication of a copy from a digitised relief & weeks (greyscale nanolithography)\\
\bottomrule
\end{tabularx}
\caption{Design estimates for the micro-fractured disc. None of these figures has been validated experimentally.}
\label{tab:disc}
\end{table}

\paragraph{Status of the properties.} B1 follows from the fuzzy extractor once
the raw error rate is confirmed. B2 holds by construction for the copies, and
for the workshop only if the master is destroyed without having been measured.
B4 rests on the absence of long-range correlation in the relief; unlike the
delay-based electrical PUFs broken in~\cite{ruhrmair}, a response here is a
function of local heights only, but this has to be established by measurement.
B5 is plausible for a mineral, sealed object and has to be confirmed by
accelerated ageing.

B3 holds in a weak form only. A legitimate reading touches a fraction of about
$10^{-6}$ of the surface, whereas predicting arbitrary responses requires
digitising all of it; but $\tau$ is counted in hours, not in years, against an
equipped adversary who takes a copy away. \emph{The protocol of
Section~\ref{sec:tfcpb} is therefore designed so that a fully digitised copy
gives nothing by itself.}

\subsection{The wish list of the second paper}

Section~4.2 of~\cite{p2} listed the properties expected from objectified shares.
Table~\ref{tab:wishes} compares them with the candidate instantiation.

\begin{table}[H]
\centering
\small
\begin{tabularx}{\linewidth}{@{}P{4.3cm}LP{3.8cm}@{}}
\toprule
Wish in \cite{p2}, Section 4.2 & Candidate instantiation & Remaining gap\\
\midrule
Clones produced without any file & direct nanoimprint from a master that is
destroyed afterwards & the workshop handles the master\\ Measurement not fixed
at production time & the challenge is chosen later, among millions of
independent ones & the reader must be specified and archived\\
Identical measurements on every copy & exact agreement through a fuzzy extractor
& raw error rate to be measured\\ Scanning should not suffice to produce a clone
& digitisation takes hours and leaves traces & scanning suffices to compute
responses\\
Stability over long periods & mineral, passive, sealed object & ageing to be
measured; readers are perishable\\ Access to the object should not be enough &
object and challenge are kept apart & --\\
\bottomrule
\end{tabularx}
\caption{The wish list of \cite{p2} and the candidate instantiation.}
\label{tab:wishes}
\end{table}

\subsection{Other uses of baryonic keys}
\label{sec:uses}

Inheritance is the use that motivated this work, but properties B1--B5 are of
interest wherever a small group of parties must share secrets that should exist
in no memory. We outline six such uses and relate them to recent work. They are
not analysed in this paper, and the caveats of Section~\ref{sec:disc} apply to
all of them. In each case the workshop plays the role that a trusted dealer
plays in the cryptographic literature, with the difference that it hands out
objects rather than digital secrets.

\begin{table}[H]
\centering
\small
\begin{tabularx}{\linewidth}{@{}P{3.6cm}LP{3.3cm}@{}}
\toprule
Use & What the key provides & Related work\\
\midrule
Post-quantum pre-shared keys between sites & the same fresh key at $n$ sites for
each epoch, stored nowhere; a compromised host extracts key material no faster
than the reading device allows & \cite{rfc8784,rfc9867}\\ Common coin for
asynchronous Byzantine agreement & identical random bits at all holders; with
one key per subset of parties, a threshold coin without distributed key
generation & \cite{rabin,cks,kms,das,fengtang,cdi}\\
Randomness beacon & responses committed by hash at setup and revealed on
schedule; values are fixed at fabrication and cannot be biased &
\cite{sokdrb,drand}\\ Threshold signatures & per-subset secrets for pseudorandom
secret sharing, re-derived at each use; deterministic nonces &
\cite{arctic,cdi}\\
Secure aggregation & pairwise masks without a key-agreement phase &
\cite{bonawitz,flamingo,opa}\\ Trusted components for replication with $2f+1$
replicas & the secret shared by the trusted components, re-derived at start-up
and never stored & \cite{minbft,gupta,bessani}\\
\bottomrule
\end{tabularx}
\caption{Uses of baryonic keys beyond inheritance.}
\label{tab:uses}
\end{table}

\paragraph{Pre-shared keys.} RFC~8784 mixes a preshared key into the key
derivation of IKEv2 so that traffic recorded today cannot be decrypted later by
a quantum computer~\cite{rfc8784}. RFC~9867, published in November 2025, extends
the mechanism to the initial exchange and to rekeying, the earlier specification
having assumed static keys~\cite{rfc9867}. Both leave open how the preshared key
is distributed and where it is kept. A baryonic key shared by two or more sites
answers both questions: the sites read the same response to a fresh public
challenge at each epoch, and the key exists in memory only while it is used.
Since responses are independent (B4), a host that was compromised for a limited
time has obtained the keys it was able to read during that time and no others.

\paragraph{Common coins.} Asynchronous Byzantine agreement needs randomness,
classically in the form of a common coin~\cite{rabin,cks}. The coin
of~\cite{cks} relies on a private setup performed by a trusted dealer, and
removing this setup is an active line of research: asynchronous distributed key
generation~\cite{kms}, consensus with neither trusted setup nor public-key
cryptography~\cite{das}, and a common coin that only requires a silent public
setup~\cite{fengtang}. A baryonic key takes the opposite route and keeps a
dealer, but a physical one. If one key is fabricated for each subset of $N-f$
parties among $N$, which is the replicated secret sharing of~\cite{cdi}, any
coalition of $f$ holders lacks one key; the exclusive-or of the responses of all
keys to a given challenge is then a coin that such a coalition cannot predict
before an honest holder reveals its values. One round of messages is needed,
since no party holds every key: each holder announces its responses, and a
response is accepted once $f+1$ holders of the same key agree on it. No
key-generation protocol is run, and no computational assumption is added to the
physical ones. For $N = 4$ and $f = 1$ this requires four keys of three copies
each.

\paragraph{Randomness beacons.} Distributed randomness beacons are expected to
be unbiasable and unpredictable~\cite{sokdrb}; deployed beacons such as
drand~\cite{drand} rely on threshold signatures. With a baryonic key, the
Donor-side technique of Section~\ref{sec:tfcpb} applies directly: commitments to
future responses are published at setup, and one response is revealed per period
and checked by hashing. The values are fixed at fabrication, so no participant
can influence them, and every key has several holders, so none can withhold a
value. The capacity of a key bounds the frequency of such a beacon.

\paragraph{Threshold signatures.} Arctic~\cite{arctic} is a deterministic and
stateless threshold Schnorr signature for small groups, built on a verifiable
pseudorandom secret sharing whose secret keys are indexed by subsets of signers,
in the manner of~\cite{cdi}. Secrets indexed by subsets are exactly what a
family of baryonic keys provides: they can be re-derived from the discs at each
signature instead of being stored, and they are distributed by the fabrication
itself rather than by an interactive ceremony.

\paragraph{Secure aggregation.} In the protocol of Bonawitz et
al.~\cite{bonawitz}, each pair of clients agrees on masks that cancel in the
sum, so that the server learns the aggregate only. Later systems reduce the cost
of the setup that establishes these pairwise secrets, by performing it once for
many rounds~\cite{flamingo} or by letting each client speak once~\cite{opa}. For
a small consortium, baryonic keys held in pairs supply the masks with no
key-agreement phase at all.

\paragraph{Trusted components.} Byzantine fault-tolerant replication needs only
$2f+1$ replicas when each replica hosts a trusted component that prevents
equivocation. In one variant of MinBFT~\cite{minbft}, the trusted components
authenticate their certificates with a secret key that all of them share.
Whether trusted components are best used to reduce the number of replicas is
debated~\cite{gupta,bessani}; in either case a baryonic key in $2f+1$ copies is
a way to provision the shared secret without a key ceremony.

%=====================================================================
\section{System and threat model}
\label{sec:model}

\subsection{Parties}

Besides the Donor, the Witnesses and the Registrars of TFCP, we consider: the
\emph{Custodians}, who each keep one sealed copy of a baryonic key; the
\emph{envelope keepers} (typically a notary), who keep sealed envelopes
containing challenges; the \emph{heirs}; the \emph{validators} of the
blockchain; the \emph{workshop} that fabricates the key; and the \emph{reading
devices}.

\subsection{Assumptions}

\begin{assumption}[Cryptography]\label{as:crypto}
$\Hash$ is a hash function with 256-bit outputs modelled as a random oracle;
$\KDF$ is a key-derivation function with the same idealisation; $\AEnc$ is an
authenticated encryption scheme.  \end{assumption}

\begin{assumption}[Blockchain]\label{as:chain}
The blockchain is safe and live. A transaction submitted with an adequate fee is
included within a bounded delay, unless validators are paid to censor it; the
cost of censorship grows with its duration~\cite{winzer,tsabary}.
\end{assumption}

\begin{assumption}[Physical key]\label{as:key}
The succession key satisfies B1--B5 with the parameters of Table~\ref{tab:disc}.
\end{assumption}

\begin{assumption}[Separation]\label{as:separation}
No party other than the Donor holds both a copy of the key and an envelope
before the Acknowledgment.  
\end{assumption}

\begin{assumption}[Availability]\label{as:avail}
After the death, a qualified set of Registrars, at least one Custodian with a
readable copy, and a reading device are available.  
\end{assumption}

\subsection{Adversaries}

We consider the following adversaries, alone or in coalition.

\begin{table}[H]
\centering
\small
\begin{tabularx}{\linewidth}{@{}P{2.1cm}L@{}}
\toprule
$\mathcal{A}_{\mathrm{out}}$ & An outsider who observes the blockchain and
submits transactions.\\ $\mathcal{A}_{\mathrm{reg}}$ & A coalition of
Registrars, possibly all of them, possibly bribed~\cite{bonneau,mccorry}.\\
$\mathcal{A}_{\mathrm{wit}}$ & A wealthy Witness, able to reach the Threshold
alone and to create any number of accounts.\\ $\mathcal{A}_{\mathrm{val}}$ &
Rational validators who accept bribes to delay or reorder
transactions~\cite{winzer,tsabary,daian}.\\
$\mathcal{A}_{\mathrm{cust}}$ & A coalition of Custodians. They have unlimited
time with their copy and may digitise it entirely.\\
$\mathcal{A}_{\mathrm{thief}}$ & A party with temporary physical access to a
copy, for a time that may exceed $\tau$.\\
$\mathcal{A}_{\mathrm{env}}$ & An envelope keeper who opens the envelopes.\\
$\mathcal{A}_{\mathrm{dev}}$ & A malicious or compromised reading device.\\
$\mathcal{A}_{\mathrm{ws}}$ & A workshop that digitises the master before
destroying it.\\
$\mathcal{A}_{\mathrm{donor}}$ & A Donor who colludes with an accomplice to
defraud Witnesses.\\ $\mathcal{A}_{\mathrm{Q}}$ & A future adversary with a
quantum computer~\cite{grover}.\\
\bottomrule
\end{tabularx}
\caption{Adversaries.}
\label{tab:adv}
\end{table}

\subsection{Security goals}

\begin{description}[style=nextline,leftmargin=1.2cm]
  \item[G1 -- Anonymity.] Before the Acknowledgment, the link between the civil
identity of the Donor and the Donor's account is hidden.  \item[G2 -- No
premature inheritance.] No inheritance transaction is executed while the Donor
is alive.
  \item[G3 -- Heir exclusivity.] The share attached to an heir is paid only to
the party to whom the Donor's arrangements give the corresponding credentials.
  \item[G4 -- Liveness.] After the death, the inheritance is eventually
executed.
  \item[G5 -- Witness fairness.] Witnesses who testify truthfully do not lose
their Ante.
  \item[G6 -- Longevity.] The above goals are not invalidated by foreseeable
cryptanalytic progress over several decades.  
\end{description}

%=====================================================================
\section{TFCP-B}
\label{sec:tfcpb}

\subsection{Principle}

The Wills are encrypted under a key derived from two halves:
\[
  K_W \;=\; \KDF\bigl(S \concat R_W \concat \mathit{id}\bigr),
\]
where $S$ is a 256-bit secret shared among the Registrars~\cite{shamir}, $R_W =
R(c_W)$ is the response of the succession key to a \emph{public} challenge
$c_W$, and $\mathit{id}$ identifies the protocol instance. The Registrars
release $S$ only after the Threshold has been reached and the Deliberation Time
has elapsed: $S$ acts as a time gate. Computing $R_W$ requires a copy of the
key: it acts as a physical confirmation. The rights of the heirs are attached to
other challenges of the same key, which are kept secret.

Security then rests on three independent locks (Table~\ref{tab:locks}), and the
third parties involved are \emph{powerless}: the notary keeps envelopes, the
bank keeps an object, the Registrars keep a secret, and none of them holds a key
or anything from which a key can be computed.

\begin{table}[H]
\centering
\small
\begin{tabularx}{\linewidth}{@{}P{1.6cm}P{3.6cm}P{3.6cm}L@{}}
\toprule
Lock & Carried by & Kept by & Useless alone because\\
\midrule
Physical & a copy of the key & Custodians & without the challenge, $2^{256}$
challenges would have to be tried\\ Digital & a 256-bit challenge in a sealed
envelope & envelope keeper or heir & without a copy, it designates nothing\\
Temporal & the Acknowledgment & Witnesses, Registrars, the network & before it,
an heir's secret confers no right\\
\bottomrule
\end{tabularx}
\caption{The three locks.}
\label{tab:locks}
\end{table}

\subsection{Setup}

\begin{enumerate}
  \item \textbf{Key.} A succession key is fabricated in $n = 3$ to 5 copies and
the master is destroyed. The Donor keeps one copy; each of the others is sealed
and handed to a Custodian.  \item \textbf{Challenges.} The Donor draws a public
challenge $c_W$, secret challenges $c_1, \dots, c_h$ uniformly in $\{0,1\}^{256}$ 
(one per heir slot), a rehearsal challenge $c_0$, and public audit challenges
$a_1, \dots, a_{50}$. For each of them the Donor runs $\Gen$ on a reading of his
own copy and obtains a response and helper data. The workshop takes no part in 
this step.  
  \item \textbf{Registrars.} They are recruited as in TFCP. The Donor draws $S$,
shares it with a low threshold, and publishes $\Hash(S)$ and a commitment to
each share.  
  \item \textbf{Wills.} The Donor publishes, under the identity of
the Security Deposit, the ciphertext $\AEnc_{K_W}(\text{link},
\text{signature})$ together with $\mathit{id}$, $c_W$, the helper data, $C_W =
\Hash(K_W)$, and the hash of all helper data of the instance.
  \item \textbf{Slots.} For each heir slot $i$, the pair $(C_i, \alpha_i)$ is
recorded, where $C_i = \Hash(R(c_i))$ and $\alpha_i$ is the share of the estate.
The audit commitments $\Hash(R(a_y))$ are recorded as well.  
  \item \textbf{Envelopes.} Each $c_i$ is sealed in an envelope labelled with the name
of an heir or with a role, and entrusted to an envelope keeper or given to the
heir.
  \item \textbf{Announcement.} It is published as in TFCP.
\end{enumerate}

\subsection{Acknowledgment}

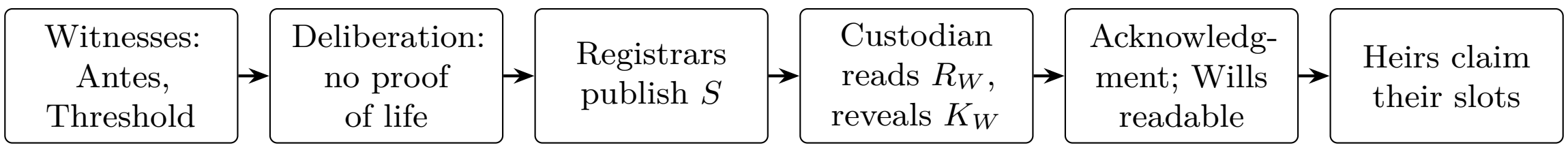
\begin{figure}[H]
\centering
\begin{tikzpicture}[node distance=3mm,
  box/.style={draw,rounded corners=2pt,align=center,minimum height=13mm,text width=20mm,font=\footnotesize},
  arr/.style={-{Stealth[length=2mm]},thick}]
\node[box] (a) {Witnesses:\\Antes, Threshold};
\node[box,right=of a] (b) {Deliberation: no proof of life};
\node[box,right=of b] (c) {Registrars publish $S$};
\node[box,right=of c] (d) {Custodian reads $R_W$, reveals $K_W$};
\node[box,right=of d] (e) {Acknowledg\-ment; Wills readable};
\node[box,right=of e] (f) {Heirs claim their slots};
\draw[arr] (a)--(b); \draw[arr] (b)--(c); \draw[arr] (c)--(d); \draw[arr] (d)--(e); \draw[arr] (e)--(f);
\end{tikzpicture}
\caption{TFCP-B after the death of the Donor.}
\end{figure}

\begin{enumerate}
  \item Witnesses transfer their Antes until the Threshold is reached. Antes are
locked until the instance terminates.  
  \item During the Deliberation Time $T_d$, a transaction signed by the Security
Deposit proves that the Donor is alive. The instance is then aborted and the
Antes remain on the Security Deposit.
  \item Otherwise the Registrars publish their shares; the network checks them
against the commitments and pays the Registrars.  
  \item \emph{Physical confirmation.} A Custodian reads $R_W$ on a copy,
computes $K_W$ and publishes it. The network checks that $\Hash(K_W) = C_W$ and
enacts the Acknowledgment. If no confirmation is published within a period $T_c
\gg T_d$, the instance is aborted and the Antes are returned, minus a fee.
  \item The Wills can now be decrypted by anybody, and the heir slots become claimable.
\end{enumerate}

The proof of life differs from that of~\cite{p1}, which requires a move on the
Donor's account. That account is hidden until the Acknowledgment, and~\cite{p1}
does not say how the network observes it. If the Registrars decrypt the Wills in
order to learn which account to watch, a false alarm discloses the link. A
signature by the Security Deposit, which is public, avoids this. The
consequences of this choice are analysed in Section~\ref{sec:prooflife}.

\subsection{Claiming a slot}
\label{sec:claim}

After the Acknowledgment, an heir obtains access to a copy and opens his
envelope. He reads $R_i = R(c_i)$ and claims in two transactions:
\begin{enumerate}
  \item \emph{commit}: publish $T = \Hash(R_i \concat \mathit{addr} \concat \mathit{nonce} \concat \mathit{id})$, where $\mathit{addr}$ is an address created for the occasion;
  \item \emph{reveal}: after $\Delta$ blocks, publish $(R_i, \mathit{addr}, \mathit{nonce})$. The contract checks that $\Hash(R_i) = C_i$ and that a matching commitment is at least $\Delta$ blocks old, and pays $\alpha_i$ to $\mathit{addr}$.
\end{enumerate}
A slot that is not claimed within a period $T_r$ follows a default rule set by
the Donor. This is a hash-locked payment in the sense of~\cite{lightning},
preceded by a commitment that prevents a party who observes the reveal from
claiming first.

The heir needs no prior key, wallet or presence on the blockchain, and the claim
involves no signature scheme other than the one protecting his newly created
address. A single key serves all heirs, since by B4 reading one slot gives no
information on the others. On a blockchain without such contracts, the slot is
replaced by a public key derived from $R_i$ and by a time-locked recovery path
in the manner of~\cite{liana}.

\subsection{Rehearsal and proof of custody}

The rehearsal slot, of small value, can be claimed while the Donor is alive. It
lets the heirs test the copy, the reading device, the envelope and the procedure
without revealing anything about the real slots. Every year $y$, each Custodian
reads $R(a_y)$ and publishes it; the network checks it against the recorded
commitment. A failing Custodian is thus detected while the Donor can still
replace him. These two mechanisms address the absence of feedback noted in
Section~4.6 of~\cite{p2}.

\section{From TFCP to TFCP-B: the changes and their rationale}
\label{sec:evolution}

This section makes explicit what TFCP-B changes with respect to TFCP, why, and
at what price. Attack numbers refer to Tables~\ref{tab:attacks}
and~\ref{tab:attacksb} of the next section.

\subsection{What is kept}

The economic core of TFCP is unchanged: the Donor and the Security Deposit, the
Witnesses and their Antes, the Threshold, the Deliberation Time, the Registrars
with their Bail and their fees, the encrypted Wills and the public Announcement.
TFCP-B remains independent of the underlying blockchain, and an instance that
never reaches its Threshold costs nothing more than in TFCP.

\subsection{The two protocols step by step}

\begin{table}[H]
\centering
\small
\begin{tabularx}{\linewidth}{@{}P{2.7cm}LL@{}}
\toprule
Step & TFCP~\cite{p1} & TFCP-B\\
\midrule
Setup: secrets & Shares of the key of the Wills are given to the Registrars &
shares of $S$ are given to the Registrars, with public commitments; copies of
the succession key are given to the Custodians\\ Setup: Wills & encrypted under
the key shared among the Registrars & encrypted under $\KDF(S \concat R_W
\concat \mathit{id})$; $\Hash(K_W)$ is published\\
Setup: heirs & not specified & slots $(C_i, \alpha_i)$ are recorded; challenges
are sealed in envelopes\\ Trigger & the Antes reach the Threshold & unchanged\\
Proof of life & a move on the Donor's account, which is hidden & a signature by
the Security Deposit, which is public\\ End of the Deliberation Time & the
Registrars decrypt the Wills; the Acknowledgment follows & the Registrars
publish $S$; a Custodian publishes $K_W$; the Acknowledgment follows; abort
after $T_c$ otherwise\\
Inheritance & transactions are triggered; their content is out of scope & heirs
claim their slots by commit and reveal\\ During the Donor's life & no feedback &
rehearsal slot; yearly proof of custody\\
\bottomrule
\end{tabularx}
\caption{TFCP and TFCP-B, step by step.}
\label{tab:steps}
\end{table}

\subsection{Seven changes}

\paragraph{C1 -- Two-factor decryption of the Wills.}
\emph{Why.} In TFCP the Registrars can pool their Shares at any time and learn
which account belongs to the Donor (attacks A1 and A2).  \emph{What.} The key of
the Wills is derived from the Registrars' secret $S$ and from a response $R_W$
of the succession key. The Registrars hold a time gate, not a decryption key.
\emph{Price.} A copy of the succession key must be readable after the death
(A12).

\paragraph{C2 -- Proof of life given by the Security Deposit.}
\emph{Why.} TFCP requires a move on an account that the network cannot identify
before the Acknowledgment; if the Registrars decrypt the Wills to identify it, a
false alarm discloses the link (A3).  \emph{What.} A transaction signed by the
Security Deposit, which is public, proves life.
\emph{Price.} The proof-of-life key is no longer the key that controls the
assets. Its theft allows a third party to abort instances after the death and to
collect the Antes (A10).

\paragraph{C3 -- Physical confirmation before the Acknowledgment.}
\emph{Why.} In TFCP, silence during the Deliberation Time is sufficient. A Donor
who is in hospital, or whose transaction is censored, is declared dead (A5, A8).
\emph{What.} After the Registrars have published $S$, a holder of a copy must
publish $K_W$. Without it the instance is aborted after $T_c$ and the Antes are
returned, minus a fee.
\emph{Effect.} The Donor is no longer obliged to watch the blockchain
permanently: an unanswered false alarm fails by default.  \emph{Price.} The
Acknowledgment takes longer and depends on a Custodian. A Custodian who is also
an heir gains no protection from this step against himself (A8).

\paragraph{C4 -- Heir slots.}
\emph{Why.} TFCP stops at the Acknowledgment. The delivery of rights to heirs,
in particular to heirs who are under age, have no wallet or are not known in
advance, was an open question of~\cite{p1,p2}.  \emph{What.} Each heir is
represented by a hash commitment; the corresponding secret is computed from a
copy of the key and a challenge kept in an envelope, and is claimed by commit
and reveal.
\emph{Price.} The separation assumption (Assumption~\ref{as:separation}) and the
trust placed in the reading device at claim time (A14, A15).

\paragraph{C5 -- Public commitments.}
\emph{Why.} A Registrar could publish a wrong Share, and public helper data
could be modified (A11, A17).  \emph{What.} Commitments to $S$, to each share
and to the helper data are recorded at setup, so that any deviation is detected.
\emph{Price.} A few additional hashes on-chain.

\paragraph{C6 -- Feedback during the Donor's life.}
\emph{Why.} In TFCP, as in any inheritance scheme, the customer is not the end
user, and nothing can be tested before the death~\cite{p2}.  \emph{What.} A
rehearsal slot lets the heirs practise the claim; a yearly proof of custody
shows that each Custodian still holds a readable copy.
\emph{Price.} One reading per Custodian and per year, and a negligible part of
the capacity of the key.

\paragraph{C7 -- Hash-based credentials.}
\emph{Why.} An inheritance may be executed decades after its setup, and the
signature schemes of today may not survive (A20).  \emph{What.} Slots,
confirmation and commitments rely on hashing with 256-bit secrets. The heirs
hold no key that could become obsolete.
\emph{Price.} Slots require a blockchain that can check hash preimages and
timing conditions; otherwise the time-lock variant of
Section~\ref{sec:discussion} is used.

\subsection{Net effect}

In TFCP, anonymity and correct execution rest on one population, the Registrars,
which is recruited on an open market and assumed to be honest but curious for
decades. TFCP-B replaces this single assumption by a conjunction: the
Registrars' secret, a copy of the succession key, and, for the funds, an
envelope. No party of a single kind can break anonymity or divert a share. The
counterpart is a set of organisational assumptions that TFCP did not need, on
Custodians, envelope keepers and reading devices, and the physical logistics
that come with them.

%=====================================================================
\section{Security analysis}
\label{sec:analysis}

We first state what TFCP-B guarantees under the assumptions of
Section~\ref{sec:model}, then review attack models from the literature. The
arguments are informal.

\subsection{Guarantees}

\begin{claim}[Anonymity against Registrars]\label{cl:reg}
Under Assumptions~\ref{as:crypto} and~\ref{as:key}, a coalition of all
Registrars and outsiders that has no access to a copy learns nothing about the
link before the Acknowledgment.  \end{claim}
\noindent\emph{Argument.} The coalition knows $S$, $c_W$ and the helper data. By
B2, $R_W$ retains close to 256 bits of entropy given the helper data, so $K_W$
is indistinguishable from random and the ciphertext hides the link.

\begin{claim}[Anonymity against Custodians]\label{cl:cust}
A coalition of all Custodians, of $\mathcal{A}_{\mathrm{thief}}$ and of
$\mathcal{A}_{\mathrm{ws}}$, holding fewer shares than the threshold, learns
nothing about the link before the Acknowledgment, even if every copy is fully
digitised.  \end{claim}
\noindent\emph{Argument.} The coalition can compute $R_W$ but has no information on $S$.

\begin{claim}[Physical confirmation]\label{cl:conf}
No Acknowledgment is enacted unless a party with access to a copy, or to a
digitised copy, takes part after the release of $S$.  \end{claim}
\noindent\emph{Argument.} Enacting requires a preimage of $C_W$, hence both $S$ and $R_W$.

\begin{claim}[Heir exclusivity]\label{cl:heir}
Under Assumptions~\ref{as:crypto}--\ref{as:separation}, and provided the reveal
transaction of the legitimate claimant is not censored for $\Delta$ blocks, slot
$i$ is paid to the holder of envelope $i$ who has access to a copy.  \end{claim}
\noindent\emph{Argument.} A valid claim requires $R_i$. Without $c_i$, even a
fully digitised copy leaves $2^{256}$ candidate challenges to test against
$C_i$. Without a copy, $c_i$ gives nothing. A party who learns $R_i$ from the
reveal must first publish its own commitment and wait $\Delta$ blocks.

\begin{claim}[No value before the Acknowledgment]\label{cl:before}
Whatever the coalition, knowledge of $R_i$ before the Acknowledgment gives no
control over the estate.  \end{claim}
\noindent\emph{Argument.} Slots are claimable only after the Acknowledgment,
which is enforced by the contract and not by secrecy.

\subsection{Attack models and resistance}

Tables~\ref{tab:attacks} and~\ref{tab:attacksb} summarise the analysis; the
following subsections discuss the cases that require more than one line.

\begin{table}[H]
\centering
\footnotesize
\begin{tabularx}{\linewidth}{@{}P{0.55cm}P{3.5cm}P{1.2cm}LL@{}}
\toprule
& Attack model & Goal & TFCP~\cite{p1} & TFCP-B\\
\midrule
A1 & Collusion of Registrars & G1 & \textbf{Broken.} Registrars can pool their
Shares at any time; acknowledged in~\cite{p1}. & \textbf{Resists}
(Claim~\ref{cl:reg}).\\ A2 & Bribery of Registrars~\cite{bonneau,mccorry} & G1 &
\textbf{Broken} as soon as the bribe exceeds the expected loss of Bail; the
collusion is undetectable. & \textbf{Resists} unless a Custodian is bribed as
well.\\
A3 & Paid false alarm to disclose the link & G1 & \textbf{Broken} if the
Registrars decrypt in order to observe the Donor's account; cost: one Threshold.
& \textbf{Resists}: $S$ alone discloses nothing.\\ 
A4 & Chain analysis of the Security Deposit~\cite{meiklejohn} & G1 &
\textbf{Outside the protocol}: the Security Deposit must be funded without an
on-chain link to the Donor. & Same.\\ 
A5 & Wealthy Witness, Donor incapacitated & G2 & \textbf{Broken}; noted
in~\cite{p1}. & \textbf{Resists} unless the adversary also has access to a copy
(Claim~\ref{cl:conf}).\\
A6 & Flash-loan funding of the Threshold~\cite{qin} & G2 & \textbf{Not
applicable} if Antes stay locked beyond one transaction. & Same.\\ 
A7 & Sybil Witnesses & G2 & \textbf{Irrelevant} for a stake-weighted Threshold;
defeats the countermeasure of~\cite{p1} based on several separate transactions.
& The physical confirmation cannot be multiplied.\\ A8 & Censorship of the proof
of life~\cite{winzer,tsabary} & G2 & \textbf{Vulnerable}: wrongful
Acknowledgment if the Donor's transaction is censored during $T_d$. &
\textbf{Partial}: the adversary also needs a copy; a malicious Custodian
qualifies.\\
A9 & Shallow account, colluding Donor & G5 & \textbf{Vulnerable}; noted
in~\cite{p1}. & \textbf{Vulnerable.}\\ 
A10 & Theft of the proof-of-life key &
G4, G5 & The key is the Donor's key: the thief can take the assets directly. &
\textbf{Vulnerable}: the thief can veto after the death and capture the Antes
(Section~\ref{sec:prooflife}).\\
A11 & Registrars withhold or corrupt Shares & G4 & Deterred by the Bail;
requires a qualified set over decades. & Same deterrent; commitments make
corruption detectable; a low threshold is affordable.\\ \bottomrule
\end{tabularx}
\caption{Attacks on the acknowledgment process, and resistance of TFCP and TFCP-B.}
\label{tab:attacks}
\end{table}

\begin{table}[H]
\centering
\footnotesize
\begin{tabularx}{\linewidth}{@{}P{0.55cm}P{3.5cm}P{1.2cm}LL@{}}
\toprule
& Attack model & Goal & TFCP~\cite{p1} & TFCP-B\\
\midrule
A12 & Loss or destruction of all copies & G4 & Not applicable. &
\textbf{Vulnerable}; mitigated by $n$ sites, yearly proof of custody and a
time-locked fallback.\\ 
A13 & Digitisation of a copy by a thief or a Custodian &
G1, G3 & Not applicable. & \textbf{Resists}
(Claims~\ref{cl:cust},~\ref{cl:heir}).\\
A14 & Envelope keeper colluding with a holder of a copy & G3 & Not applicable. &
\textbf{Vulnerable} after the Acknowledgment; nothing before
(Claim~\ref{cl:before}).\\ 
A15 & Malicious reading device & G3 & Not applicable.
& \textbf{Vulnerable} at claim time (Section~\ref{sec:device}).\\
A16 & Front-running of a claim~\cite{daian,eskandari} & G3 & Not specified. &
\textbf{Resists} up to censorship for $\Delta$ blocks.\\ 
A17 & Helper-data manipulation~\cite{dkrs} & G3, G4 & Not applicable. & 
\textbf{Resists}: helper data are bound by an on-chain hash.\\
A18 & Modelling attack on the key~\cite{ruhrmair} & G1, G3 & Not applicable. & 
Resists \textbf{if B4 holds}; to be validated.\\
A19 & Malicious workshop & G1, G3 & Not applicable. & \textbf{Resists}: no more
than a Custodian.\\ 
A20 & Quantum adversary~\cite{grover} & G6 & Signatures and
Share distribution are exposed. & Slots and Wills rely on hashing with 256-bit
secrets; the proof-of-life signature must be migrated by the living Donor.\\
\bottomrule
\end{tabularx}
\caption{Attacks on custody, claims and the physical key, and long-term attacks.}
\label{tab:attacksb}
\end{table}

\subsection{Collusion, bribery and false alarms (A1--A3, A5)}

In TFCP the Shares held by the Registrars are sufficient to decrypt the Wills.
The anonymity property is therefore stated in~\cite{p1} under an
honest-but-curious assumption. Registrars are recruited on an open market and
serve for decades; pooling Shares leaves no trace, and the literature on bribery
shows that such collusions can be organised and paid for
trustlessly~\cite{bonneau,mccorry}. We consider that G1 does not hold for TFCP
against $\mathcal{A}_{\mathrm{reg}}$.

In TFCP-B the same coalition obtains $S$ and nothing else (Claim~\ref{cl:reg}).
A further consequence is that the task of a Registrar can be transferred, and
$S$ re-shared, without risk for G1. The availability of Registrars over decades,
an open question of~\cite{p1}, can then be addressed by recruiting many of them
with a low threshold.

A false alarm, in which $\mathcal{A}_{\mathrm{wit}}$ reaches the Threshold
although the Donor is alive, has two possible purposes. The first is to force
the disclosure of the link (A3); in TFCP-B, even if the Donor fails to react and
the Registrars release $S$, nothing is disclosed. The second is to obtain a
wrongful Acknowledgment while the Donor is unable to react (A5), which~\cite{p1}
identifies as dangerous. In TFCP-B the Custodians, who are close to the Donor,
simply do not confirm; the instance is aborted after $T_c$. The Donor is no
longer required to watch the blockchain permanently, which was the main drawback
of dead man's switches. The residual case is an adversary who combines the
Threshold, an incapacitated Donor and access to a copy.

\subsection{Flash loans and Sybil identities (A6, A7)}

The original paper mentions a ``flash loan attacker'' who would reach the
Threshold alone. A flash loan is valid within a single transaction and must be
repaid by the end of that transaction~\cite{qin}. An Ante that remains locked on
the Security Deposit for the Deliberation Time cannot be financed in this way,
so this attack does not apply, provided that the implementation never refunds an
Ante within the transaction that deposits it. What remains is the wealthy
Witness. The countermeasure suggested in~\cite{p1}, namely to require several
separate transactions, is ineffective against an adversary who creates several
accounts, unless the transactions are also spread over time, as~\cite{p1} notes.
In TFCP-B the second factor is physical and exists in $n$ copies only.

\subsection{Censorship during the Deliberation Time (A8)}

Both protocols rely on a time-out: the absence of a proof of life during $T_d$
is interpreted as death. Time-outs of this kind are known to be exposed to
incentive manipulation. Winzer et al.~\cite{winzer} show that validators can be
bribed through a smart contract to ignore a transaction until a time-out
elapses, and Tsabary et al.~\cite{tsabary} show that such attacks on hashed
time-locked contracts are cheaper than previously thought. Here the party with
an interest in censoring is whoever benefits from a premature Acknowledgment, in
the first place a malicious heir.

TFCP has no other defence than the length of $T_d$ and the fee that the Donor is
ready to pay. TFCP-B adds the physical confirmation, but this is of limited help
when the malicious heir is also a Custodian. We therefore classify the
resistance as partial, and recommend (i) a Deliberation Time counted in weeks,
since the cost of censorship grows with its duration; (ii) Custodians who have
no interest in the estate; and (iii) for large estates, a threshold confirmation
in which each Custodian $j$ is given a personal confirmation challenge and $R_W$
is replaced by a secret shared among the corresponding responses, so that $k$
distinct Custodians must take part.

\subsection{The proof of life and the shallow-account attack (A9, A10)}
\label{sec:prooflife}

In the shallow-account attack of~\cite{p1}, the Donor gives an accomplice the
means to prove life. After the death, the accomplice aborts the instance and
takes the Antes. No design of the proof of life can prevent this: any secret
that proves life can be handed over, and a baryonic key is no exception, since
an object does not prove who holds it. Both protocols are vulnerable. The only
limits are those mentioned in~\cite{p1}: the reputation of a named Donor, and
the pricing of this risk by Witnesses.

The choice of the key that proves life nevertheless matters. In TFCP it is the
key of the Donor's account. Delegating it amounts to giving away the assets, and
its theft is no worse than the theft of the assets. In TFCP-B it is the key of
the Security Deposit. This protects anonymity (attack A3), but it creates a key
whose theft has a specific effect: after the death of the Donor, the thief can
abort every instance and collect the Antes, which violates G4 and G5. We see two
mitigations. The key of the Security Deposit can be derived from the succession
key and a challenge known to the Donor only, so that it is not stored anywhere.
And the time-locked fallback path of Section~\ref{sec:claim} bounds the damage,
since it does not depend on the Acknowledgment.

\subsection{Custodians, envelopes and reading devices (A13--A15)}
\label{sec:device}

The design assumes that Custodians, who keep a copy for years, may digitise it
entirely: this gives $R_W$, which is useless without $S$, and no $R_i$
(Claims~\ref{cl:cust} and~\ref{cl:heir}). The weak form of B3 is therefore
sufficient for G1 and G3.

Assumption~\ref{as:separation} is organisational, and its violation is the main
residual risk for G3. An envelope keeper who opens envelope $i$ and gains access
to a copy, or to a digitised copy, can compute $R_i$. By Claim~\ref{cl:before}
this is harmless before the Acknowledgment; afterwards the coalition can claim
the slot before the legitimate heir. This is the counterpart of the trust placed
in the executor of a will, restricted to a diversion after the death. It can be
reduced by splitting each challenge between two keepers, and by giving adult
heirs their envelope directly.

A reading device sees both the challenge and the response. A malicious device
used by an heir can therefore leak $R_i$ before the heir has committed, which
defeats the commit--reveal step. The device used for claims must be trusted at
that moment. We recommend an open design without network interface, and the
possibility for an heir to use a device of his own.

\subsection{Attacks on the physical key (A17--A19)}

An adversary who modifies the public helper data can cause $\Rep$ to fail or,
for some constructions, to leak information~\cite{dkrs}. In TFCP-B the hash of
all helper data is recorded on-chain at setup, so that modified helper data are
rejected.

Modelling attacks~\cite{ruhrmair} exploit the structure of a PUF to predict
unseen responses. The relief of a micro-fractured surface has no such structure
by design, and B4 states that responses at disjoint sites are independent; an
adversary who has scanned part of a copy then knows exactly the responses that
fall in that part. B4 is an assumption until measured.

The workshop could measure the master before destroying it. In TFCP-B all
responses are generated by the Donor, so the workshop obtains at most what a
Custodian with a digitised copy obtains. Combining keys from independent
workshops would reduce this assumption further.

\subsection{Longevity (A20)}

Grover's algorithm~\cite{grover} reduces the cost of an exhaustive search over
$N$ candidates to about $\sqrt{N}$ evaluations. With 256-bit challenges and
256-bit responses, searching for a challenge that matches a public commitment,
on a digitised copy, costs about $2^{128}$ quantum evaluations. The slots, the
Wills and the confirmation rely on hashing and symmetric encryption only. The
signature of the Security Deposit depends on the signature scheme of the
blockchain; a migration can be carried out by the Donor while alive. No action
is required from the heirs, who hold no key.

\subsection{Summary}

\begin{table}[H]
\centering
\small
\begin{tabularx}{\linewidth}{@{}P{3.4cm}LL@{}}
\toprule
Goal & TFCP & TFCP-B\\
\midrule
G1 Anonymity & only against honest-but-curious Registrars & holds unless
Registrars \emph{and} a holder of a copy collude\\ 
G2 No premature inheritance &
fails if the Donor cannot react or is censored & requires in addition access to
a copy\\
G3 Heir exclusivity & not addressed & holds under Assumption~\ref{as:separation}
and with a trusted reading device\\
G4 Liveness & Registrars & Registrars, one Custodian, one device; fallback path\\
G5 Witness fairness & fails against a colluding Donor & same; also fails if the 
Security Deposit key is stolen\\
G6 Longevity & not addressed & hash-based for heirs; signature migration by the Donor\\
\bottomrule
\end{tabularx}
\caption{Security goals.}
\end{table}

%=====================================================================
\section{Discussion}
\label{sec:discussion}

\subsection{The open questions of the earlier papers}

\paragraph{Secrets sent to the future (\cite{p2}, Section 4.2).} A slot is
defined by a commitment, and the corresponding credentials are an object and a
number. The recipient need not exist, be of age, or be known when the slot is
created.

\paragraph{Flexibility of wills (\cite{p2}, Section 4.3).} The second paper
opposed the rigidity of smart contracts to unforeseen events: an heir who dies
first, a collective accident, a change in the law. With slots, the contract
fixes shares and names nobody. If the intended heir is no longer there, the
envelope follows the ordinary rules of succession and whoever receives it claims
the slot. Flexibility is thus returned to the legal system without giving it
custody of the funds. Amounts remain fixed, and may conflict with
forced-heirship rules~\cite{carata}.

\paragraph{Civil status (\cite{p2}, Section 4.4).} The heir needs no on-chain
identity. The link between a person and an envelope is established by the
envelope keeper, as it would be for any bequeathed object.

\paragraph{Long term (\cite{p2}, Section 4.5).} The key is passive and its
responses are seeds that depend on no particular algorithm. The reading device
is the perishable element; its specification has to be open and archived.

\paragraph{Feedback (\cite{p2}, Section 4.6).} The rehearsal slot and the yearly
proof of custody provide feedback during the Donor's life.

\paragraph{Atomicity (\cite{p2}, Section 4.7).} A single key can carry slots for
several blockchains. Moreover, $C_W$ can be recorded on several blockchains, the
publication of $K_W$ on one of them serving as a death signal on the others, as
the preimage does in an atomic swap. On a blockchain where no Ante is at stake,
Registrars colluding with a Custodian could raise this signal cheaply; a local
Deliberation Time remains necessary there.

\paragraph{Registrars (\cite{p1}, Section 5.1).} Their task can be transferred
and their number increased without consequence for anonymity. Whether a market
for Registrars is needed is unchanged.

\paragraph{Minors and unknown heirs (\cite{p1}, Section 5.1).} The copy waits in
a vault and the envelope at the notary's. Neither the guardian nor the bank can
spend alone. Envelopes can be labelled with roles, and a residual slot can cover
unforeseen cases.

\subsection{Variants}

\begin{table}[H]
\centering
\small
\begin{tabularx}{\linewidth}{@{}P{3.3cm}LL@{}}
\toprule
Variant & Principle & Use\\
\midrule
Time lock only & no distributed Acknowledgment; the heir's key, derived from the
copy, becomes valid after a period of inactivity & blockchains without the
required contracts\\ No Registrars & $K_W = \KDF(R_W \concat \mathit{id})$;
Custodians can read the Wills & small estates\\
Registrars with keys & $S$ is derived from baryonic keys held by the Registrars:
no digital Share at rest & few institutional Registrars\\ Threshold confirmation
& $k$ distinct Custodians must confirm & large estates (attack A8)\\
\bottomrule
\end{tabularx}
\caption{Variants of TFCP-B.}
\end{table}

%=====================================================================
\section{Costs}
\label{sec:costs}

Table~\ref{tab:market} gives the prices of existing services and components that we could verify at the time of writing.

\begin{table}[H]
\centering
\small
\begin{tabularx}{\linewidth}{@{}P{6.2cm}P{4.2cm}L@{}}
\toprule
Item & Price & Source\\
\midrule
Casa Standard (3-key vault, inheritance) & USD 250 per year & \cite{casa}\\
Casa Premium (5-key vault) & USD 2,100 per year & \cite{casa}\\
Casa Private Client & USD 7,500 per year & \cite{casa}\\
Nunchuk, plan with inheritance & USD 480 per year & as reported in~\cite{spark}\\
Liana (software) & free, open source & \cite{liana}\\
Optical pickup unit (CD/DVD/Blu-ray) & about USD 10 per piece & \cite{sekar}\\
\bottomrule
\end{tabularx}
\caption{Verified market prices (November 2026).}
\label{tab:market}
\end{table}

For the baryonic key itself there is no market price, and the following figures
are our estimates. The optical head is a commodity (Table~\ref{tab:market}), and
polished borosilicate substrates of 100\,mm are sold by laboratory suppliers for
a few tens of dollars per unit; the cost is dominated by the replication,
coating and sealing steps, performed in a small trusted facility. We estimate a
set of three copies at EUR 95--140 and a set of five at EUR 140--220 in small
series, a prototype reading device at EUR 2,000--5,000, and a device produced in
series at EUR 150--400. A device kept by a notary or a bank would serve all
their clients.

\begin{table}[H]
\centering
\small
\begin{tabularx}{\linewidth}{@{}P{6.6cm}L@{}}
\toprule
256-bit response, 8\,\% error margin & 122 sites, read in about 35\,s\\
Kit for 11 heirs & 63 responses: 1 for $R_W$, 11 slots, 1 rehearsal, 50 audits\\
Surface used & 7,686 sites, i.e.\ about 0.01\,\% of one side\\
Readings at setup & about 37 minutes\\
Claim by an heir & one reading, two transactions\\
\bottomrule
\end{tabularx}
\caption{Sizing of a succession kit under the estimates of Table~\ref{tab:disc}.}
\end{table}

Over thirty years, the least expensive assisted service of
Table~\ref{tab:market} amounts to USD 7,500, and it depends on the continued
existence of its provider. A baryonic key is a one-off expense, but it requires
access to a reading device when the time comes.

%=====================================================================
\section{Limitations}
\label{sec:limits}

\begin{itemize}
  \item \textbf{No experimental validation.} The candidate instantiation has not
been built. The raw error rate between copies, the entropy per site, the
independence of sites, ageing, and the time needed to digitise a copy are
estimates. If $\tau$ turned out to be much shorter, the analysis of
Section~\ref{sec:analysis} would still hold, since it assumes that copies may be
digitised; if the error rate were much higher, capacity and reading time would
degrade.  
  \item \textbf{Informal analysis.} We have given arguments, not proofs.
A formal model of the protocol, including the game-theoretic behaviour of
Witnesses, Registrars and validators, remains to be written. The parameters
$T_d$, $T_c$, $T_r$, $\Delta$, the Threshold, the Antes and the Bails have to be
set, and as noted in~\cite{p1} only practice will settle some of them.
  \item \textbf{Attacks that remain.} The shallow-account attack (A9),
censorship during the Deliberation Time (A8), the theft of the proof-of-life key
(A10), the violation of the separation assumption (A14) and a malicious reading
device (A15) are not solved. 
  \item \textbf{The oracle does not prove death.}
The Acknowledgment remains an economic bet completed by a physical gesture. It
also makes a payment conditional on a death, so the fees paid to Witnesses must
remain small enough not to create an incentive to cause it.
  \item \textbf{Availability.} Enough Registrars, one readable copy and one
device are needed after the death.  
  \item \textbf{Re-centralisation.} Vaults and
notaries are back in the loop. They have no power over the funds, but they can
fail in their custody.
  \item \textbf{Law.} Jurisdiction, taxation and forced heirship are not addressed.
\end{itemize}

%=====================================================================
\section{Conclusion}

The second paper of this series suggested that the materiality of an object
could work around the perfect and non-destructive copy that digital technologies
allow. Baryonic keys make this suggestion precise: keys cloned a small, fixed
number of times, whose bits are never stored. The instantiation that we outlined
makes digitisation slow and visible rather than impossible, and the protocol was
designed accordingly, so that a digitised copy gives nothing in isolation.

Applied to the Tales From the Crypt Protocol, the primitive removes the
dependence of anonymity on the honesty of Registrars, blocks the inheritance of
an incapacitated Donor after a false alarm, and provides what the protocol had
left out of scope: a way to deliver rights to heirs who hold no key, are under
age, or are not yet known. The analysis also shows what it does not provide. The
shallow-account attack is untouched, censorship during the Deliberation Time
remains a concern, and custodians, envelope keepers and reading devices bring
assumptions of their own.

Future work is experimental, on the physical key, and formal, on the protocol.
We believe that keys which exist in a few places and nowhere else are of
interest beyond inheritance; the uses outlined in Section~\ref{sec:uses} deserve
an analysis of their own.

%=====================================================================

\end{document}